# An efficient photoelectric X-ray Polarimeter for the study of Black Holes and Neutron Stars

Enrico Costa*, Paolo Soffitta*, Ronaldo Bellazzini$^§$, Alessandro Brez$^§$, Nicholas Lumb$^§$, Gloria Spandre$^§$

*Istituto di Astrofisica Spaziale del CNR, Via Fosso del Cavaliere 100, I-00133, Rome, Italy.

$^§$Istituto Nazionale di Fisica Nucleare-Sezione di Pisa, Via Livornese 1291, I-56010 San Piero a Grado, Pisa, Italy.

**In astronomy there are basically four kinds of observations to extract the information carried by electromagnetic radiation: photometry, imaging, spectroscopy and polarimetry. By optimal exploitation of the first three techniques, X-ray astronomy has been able to unveil the violent world of compact high energy sources. Here we report on a new instrument that brings high efficiency also to X-ray polarimetry, the last unexplored field of X-ray astronomy. It will then be possible to resolve the internal structures of compact sources which otherwise would remain inaccessible, even to X-ray interferometry[1]. Polarimetry could provide a direct, visual picture of the state of matter under extreme magnetic and gravitational fields by measuring the radiation polarized through interaction with the highly asymmetric matter distribution[2,3] (accretion disk) and with the magnetic field[4,5,6,7].**

The new instrument derives the polarization information from the track of the photoelectrons imaged by a finely subdivided gas detector. Its great improvement of sensitivity (at least two orders of magnitude) will allow direct exploration of the most dramatic objects of the X-ray sky.

In binary pulsars we can directly "see" the rotation of the magnetic field around the spin axis as a swing of the polarization plane with the pulse phase. The correlation of polarization and luminosity will determine, at last, whether the emission is in the form of a *fan* or a *pencil* beam[8,7].

In accreting binaries we expect a unique signature of a harbored Black Hole: the polarization angle will be twisted in our frame by the gravitational field and this will be a continuous function of the energy [9,10,11,12].

The main breakthrough of the proposed polarimeter is, however, its capability of investigating Active Galactic Nuclei for which crucial polarization measurements have been suggested. We can separate synchrotron X-rays from Jets (Blazars)[13,14] from the emission scattered by the disk corona or by a thick torus (Seyfert-2). The effects of relativistic motions and of the gravitational field of a central Black Hole have been likely detected by iron line spectroscopy on the Seyfert-1 galaxy MCG-6-30-15[15] but this feature is not ubiquitous in AGNs. Polarimetry of the X-ray continuum provides a more general tool to explore the structure of emitting regions[16,17], to track instabilities and to derive direct information on mass and angular momentum[12] of supermassive Black Holes.

Notwithstanding this wealth of expectations, the important but unique result is, until now, the measurement, by the Bragg technique, of the polarization of the Crab Nebula[18,19]. The Stellar X-ray Polarimeter[20] (SXRP) represents the state of the art for conventional methods based on Bragg diffraction and Thomson scattering. However, Bragg polarimetry[21] is dispersive (one angle at one time) and very narrow band. Thomson polarimetry[22] is not imaging and band-limited (>5keV). This limits the sensitivity of SXRP to a few bright, galactic sources only.

The photoelectric effect is very sensitive to polarization. The electron is ejected from an inner shell with a kinetic energy which is the difference between the photon energy and the binding energy. The direction of emission is not uniform but is peaked around that of the electric field of the photons (see fig.1a). This photoelectron interacts with the surrounding matter: it is slowed by ionizing collisions with atomic electrons and scattered by coulomb diffusion on the nuclei (see fig.1b) and eventually stopped. The photoelectron leaves in the absorber a string of electron/ion pairs, marking the path from its creation to the stopping point. We call this cluster a "track": in the initial part of this track resides the information on the original electron direction and thence the key to derive the polarization of the photon. This dependence is preserved if the track is projected onto a plane perpendicular to the radiation.

In a subdivided detector extended tracks may produce coincident signals in two contiguous cells. If the radiation is polarized the orientation of these pairs is asymmetric. One can exploit this by counting coincidences in neighboring wires of proportional counters[23] or CCD pixels[24,25]. But since the detector cell is typically much larger than the electron range, the asymmetry effect strongly depends on the absorption point. This can be avoided if the cell is so small that the track is split into several cells. The first finely subdivided, self triggered device[26] was a micro-gap[27] chamber filled with a Neon-based gas mixture at 1 atmosphere. The persistent need to rotate the instrument and a still moderate modulation factor make this device a step forward, but not yet a real breakthrough.

Other instruments image on a CCD the bright track made in a gas scintillation detector. One of the two practical implementations[28] works only above 40 keV, the other[29] only at low pressure. Both use Argon which, because of the high energy of the isotropic Auger electron and larger multiple scattering, is a gas suitable for photon energies higher than the practical range for X-Ray optics.

Position sensitive gas detectors typically yield the centroid of the charge cloud, whose extent is the ultimate limit to the space resolution, a sort of noise to be kept as small as possible. We reverse this approach, trying to resolve the track to measure the interaction point and the prime direction of the photoelectron. To this end we have developed the micro-pattern gas chamber (MPGC). It consists (fig. 2, Tab.1) of a gas cell with a thick detection/drift region, a thin gas electron multiplier (GEM[30]) and a multi pixel, true two-dimensional, read-out anode. The large number of fired pixels per track allows for good track reconstruction. Additionally, the MPGC measures the energy lost in each pixel, a quantity directly related to the kinetic energy of the electron.

We filled the MPGC with a 1-atm mixture of Ne (80%)-DME (20%). Fig 3.a shows the image of a real MPGC track. An initial straighter part, with low ionization density, which carries most of the information on the starting direction (and thence on the polarization) evolves into a *skein* with high ionization density and a completely random path. Most of the photoelectron energy is lost at some distance from the initial interaction point. To verify this interpretation we let impinge on the detector, through a very thin diaphragm, photons of 5.4 keV from an unpolarized source. The loci of the centroids of each track are displaced from the interaction point and located on a circular region around it, indicating that the tracks have, and retain, a significant elongation and energy loss asymmetry ( figure 3.b, top). From each track we reconstructed the emission angle and we built a histogram. In the case of unpolarized X-ray photons of 5.4 and 5.9 keV, all the emission angles have the same probability and the histogram is flat (figure 4.b, top). When we irradiated the detector with an extended, nearly 100 % polarized source of 5.4 keV, we found a strong angular modulation (44%) that is well modeled by the expected distribution (figure 3.b, bottom), taking into account the theoretical distribution of the photoelectron and the smearing due to scattering.

An important issue is the choice of the gas filling. We used a Ne based mixture because in the energy band of interest the photoelectron track is longer and fires several pixels, while retaining reasonable efficiency. A low Z gas is less efficient to primary photon detection. However, the scattering/slowing ratio is lower as well: the track is more straight and the direction of emission can be measured more precisely. Moreover the K-edge energy is so low (0.87 keV) that the accompanying isotropic Auger electron does not blur the information on the polarization while it helps, instead, in the identification of the impact point. The use of even lower K-edge converters together with a very fine pixel size could mak low energy polarimeters (0.5-2 keV) conceivable.

With our prototype we have demonstrated the practical feasibility of a new generation of photoelectric polarimeters in the 2-10 keV band. The device can also do simultaneously good imaging (50-100 μm), moderate spectroscopy (16% fwhm at 5.4 keV), and fast, high rate timing down to 150 eV. Moreover, being truly 2-D, it is non-dispersive and does not require rotation.

We also tested our capability to model the polarization detection processes. Since absorption, slowing down, scattering and transverse diffusion of electrons in the drift are well known quantities, we may reliably predict the performance of another detector configuration better optimized for astrophysical applications. It is based on an already existing[31] VLSI readout chip while all other features are well established detector technology. We can derive the polarimetric sensitivity of such detectors when installed at the focus of a real X-ray telescope. In Tab.1 we compare the sensitivity of the present and final configuration of the MPGC with SXRP.

The MPGC requires integration periods ~100 times shorter to detect the same polarization in bright sources. With integrations of the order of one day we could perform polarimetry of Active Galactic Nuclei at the per cent level, a breakthrough in this fascinating window of high energy astrophysics. With the planned XEUS[32] telescope, polarimetry could become a new high throughput branch of X-ray astronomy.

This work is partially supported by Italian Space Agency.

**<Correspondence and request for materials should be addressed to E. Costa (e-mail: costa@ias.rm.cnr.it>**

Fig 1. Panel a (left) Basic physics of the photoelectric effect. The photoelectron is ejected in directions that carry a significant memory of the electric field of the photon. When the beam is linearly polarized the electrons are ejected preferentially around the electric field. The cross section for *s* electrons is:

$$\frac{\partial \sigma}{\partial \Omega} = r_0^2 \frac{Z^5}{137^4} \left(\frac{mc^2}{h\nu}\right)^{7/2} \frac{4\sqrt{2} \sin^2(\theta) \cos^2(\varphi)}{(1 - \beta \cos(\theta))^4},$$

where $r_o$ is the classical electron radius, Z is the atomic number of the target material and $\beta$ is the electron velocity, as a fraction of the speed of light c. The figure shows the meaning of the emission angle $\theta$ and azimuth angle $\varphi$.

Panel b (right) A simulated 6-keV photoelectron and Auger track, showing propagation in the gas and collection on a plane with sensitivity to both energy and position. The photoelectron is slowed by ionizing collisions with outer electrons of the atoms of the medium. The energy loss increases with decreasing kinetic energy (Bethe law for low energy).

$$\frac{\partial E}{\partial x} \propto \frac{1}{\beta^2} \propto \frac{1}{E_{kin}}$$

Electrons are also scattered by charges in the nuclei with no significant energy loss. This follows the screened Rutherford law:

$$\frac{\partial \sigma}{\partial \Omega} \propto \frac{Z^2 / E_{kin}^2}{(\sin^2(\vartheta/2) + \alpha_{screen})^2}$$

While scattering crucially depends on the atomic number, slowing down is only moderately dependent. The primary ionizations are then projected onto the sense plane. The charge density in each pixel is proportional to the energy loss, and, therefore related to the electron kinetic energy. The starting path has a lower charge density but it is close to the initial direction of the photoelectron and thence closely related to the photon polarization direction. The latter path has a higher charge density but it is randomized with no information on the polarization direction.

Figure 2. The micro-pattern gas detector. The photon is absorbed at some point in the drift gap. The photoelectron track is drifted by the electric field to the gas electron multiplier. This device is made of a thin (50 $\mu$m) polyimide foil perforated by many microscopic holes, where a high electric field provides the charge amplification. Finally, the charge is collected by the pixels of the MPGC anode, each one connected to an independent electronics chain. On receiving a trigger from the GEM, all the signals are analog to digital converted, so that we have the image of the track projected on the detector plane.

Figure 3. Panel a (left) Image of a real photoelectron track detected by a MPGC filled with Neon 80% and Dimethyl-Ether (20 %). Scale unit is pixel number and larger boxes correspond to a larger energy loss. Each pixel is 200 $\mu$m wide. It is possible to recognize the beginning of the track with the Auger electron followed by the weaker ionization loss of the photoelectron (top) and the end of the track with a much larger energy loss (70 % of the total charge for this specific event, bottom). For each photoelectron it is, therefore, possible to measure the initial direction of the track, which carries memory of the polarization. By reconstructing the impact point of the photon, the real position resolution is much better than that imposed by the track extension and is only a factor of two-three worse than a CCD.

Figure 3. Panel b (right). Histogram of the emission angles of the photoelectron in the detector plane as reconstructed from data like the one in figure 3.a. The top histogram is for unpolarized photons from a $Fe^{55}$ source. No preference in the track direction results in a histogram which is consistent with a flat curve. In the top figure the loci of the baricenters for a 5.4-keV pencil beam of unpolarized photons are displayed, showing how tracks retain their energy loss asymmetry.

Finally, the bottom histogram is for nearly 100 % polarized photons from a 5.4-keV extended source. The amplitude of the $\cos^2$ fit to the histogram of counts is directly related to the sensitivity of a real polarimeter. The angular phase is the direction of polarization of the incoming photons. The so called "modulation factor" (Cmax-Cmin)/(Cmax+Cmin) is measured to be 0.44. The $\chi^2_{red}$ to the fit is 1.02, a = 36.98±1.84; b=58.43±3.57; $\phi_o$=0.45°±1.73°.

|  | Present Prototype (2-10 keV) | Improved configuration (3.5-10 keV) |
|---|---|---|
| Drift/Absorption Gap | 6 mm | 30 mm |
| Drift field | 3000 V/cm | 1500 V/cm |
| Gas filling and pressure | (Ne 80% - DME 20%) 1 Atm | ( Ne 40%-DME 60%) 4 Atm |
| Gas Gain | 5000 | 2500 |
| Transverse diffusion in drift | 80 μm | <100 μm |
| GEM thickness | 50 μm copper clad kapton foil | 50 μm copper clad kapton foil |
| GEM hole geometry | 40 μm diameter 60 μm pitch | 40 μm diameter 60 μm pitch |
| GEM Voltage | 400 V | 600 V |
| Detection efficiency at 5.4 keV | 3.8 % | 91 % |
| Read-out pixel size | 200 μm | 50 μm |
| Number of pixels | 512 | 40000 |
| Read-out plane technology | Multilayer advanced PCB | VLSI |
| Track lenght/pixel size (6 keV) | 6 | 6 |
| Sensitivity to Her X1 | T = 400 s ; MDP = 10 % | T = 20 s ; MDP = 10 % |
| Sensitivity to 3C-273 | T = 2.2e$^5$ s ; MDP = 2 % | T= 4e$^4$ s ; MDP = 1 % |
| Sensitivity to MCG-6-30-15 | T = 5e$^5$ s ; MDP = 2% | T = 1e$^5$ s ; MDP = 1 % |
| Gain in the integration time over SXRP (strong sources) | 5 (over Thomson) 15 (over Bragg) | 100 (over Thomson) 300 (over Bragg) |
| Gain in the integration time over SXRP (faint sources) | 250 (over Thomson) 100 (over Bragg) | 5000 (over Thomson) 2000 (over Bragg) |

Table 1. Physical characteristics of the micro-pattern detector in the configuration now under test in our laboratory and in the improved configuration.

We also show the observing time needed to measure at 99 % confidence the shown degree of polarization (minimum detectable polarization or MDP), of a selected sample of astrophysical sources if the MPGC is placed at the focus of the SODART[33] telescope on board of Spectrum X-gamma mission. The relevant formalism can be found in Soffitta et al.[26] . For the improved design we only use photons above 3.5 keV to compute the sensitivity. Below this energy, at high pressure, the transverse diffusion and the unfavorable track length to pixel size ratio significantly reduces the modulation. Her X-1 is a Galactic binary X-ray source with a magnetic field of ~ $3.5 \cdot 10^{12}$ G measured by the detection of a cyclotron line. A high degree of linear polarization is expected and, by polarimetry, a direct measure of the angle between the magnetic field and the rotation axis can be performed with $1°$ accuracy.

3C-273 is the brightest X-ray radio-loud quasar and MCG-6-30-15 is a Seyfert 1 galaxy for which a broad (~ 80.000 km/s) iron line was observed, possibly skewed by gravitational effects[15].

Finally we compare the performances of MPGC with SXRP which is the most sensitive polarimeter foreseen in a future X-ray mission, ready to be installed in the focal plane of the SODART telescope. It simultaneously exploits the Bragg diffraction at $45°$ and the azimuth dependence on polarization of Thomson scattering. The ensemble analyzer/detector of the two stages is rotated to search for a modulation in the counting rate and to compensate systematic effects.

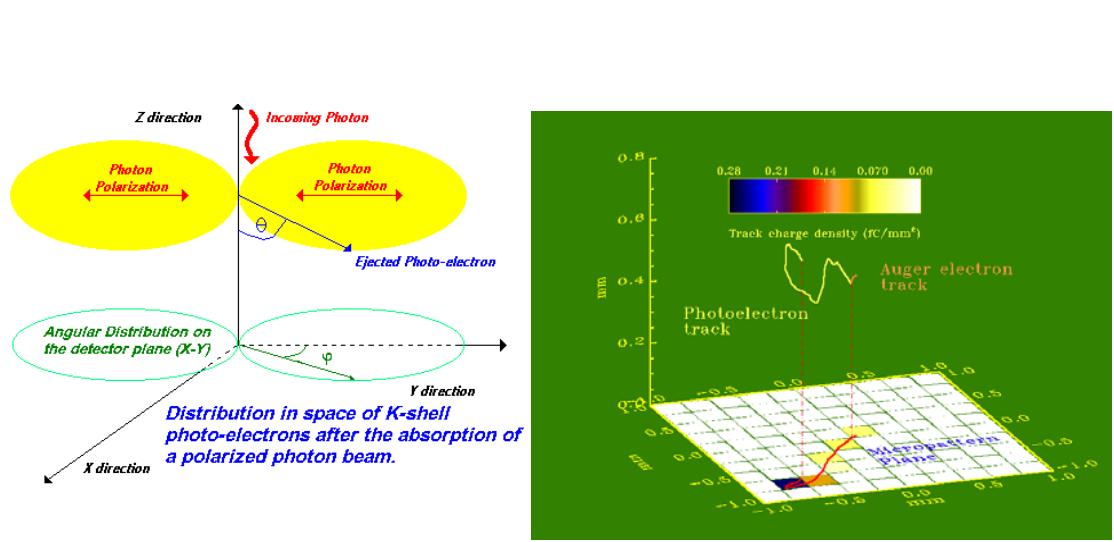

Fig. 1
Costa E.

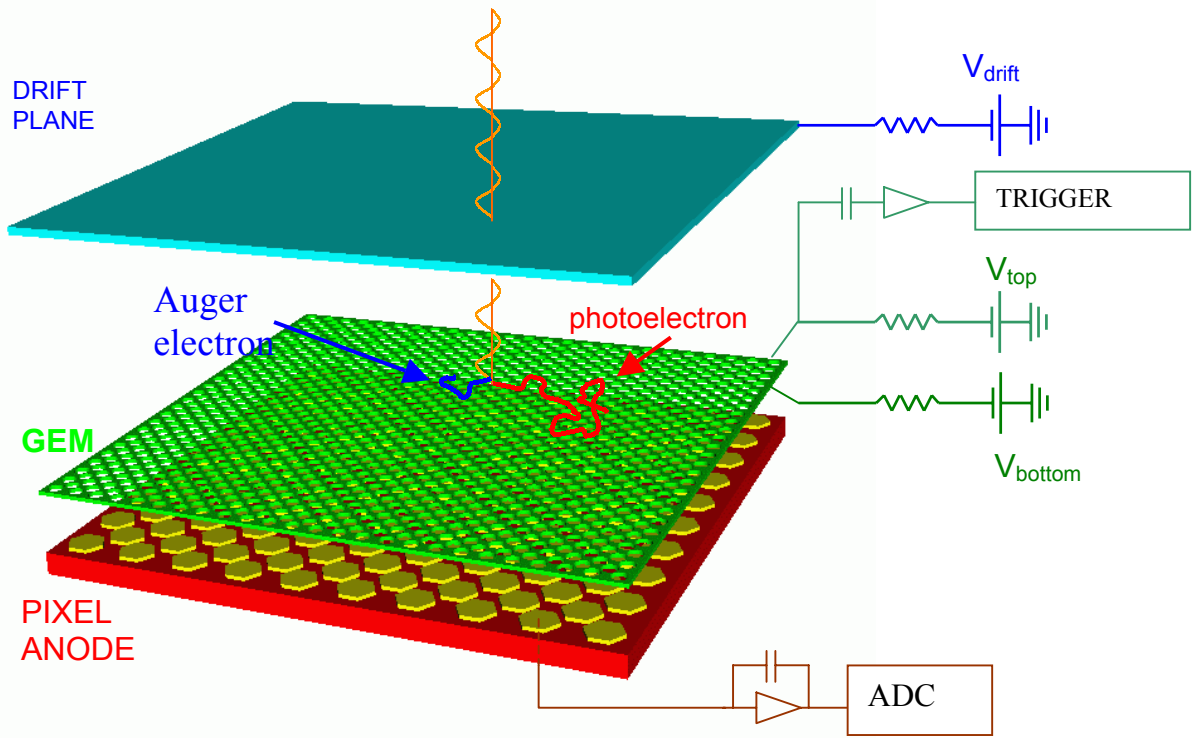

Fig. 2
Costa E.

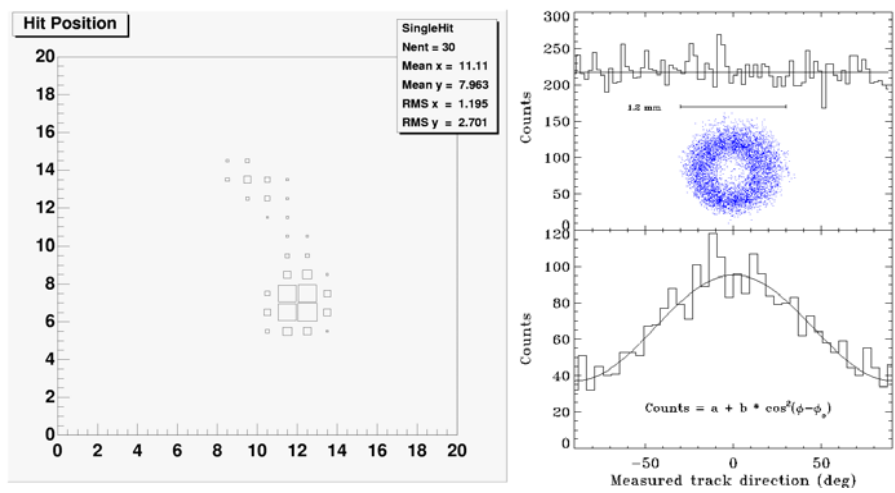

Fig. 3
Costa E.